# A MEASUREMENT OF THE EARTH'S RADIUS FOR HIGH SCHOOL STUDENTS

**Davide Neri**
**Liceo A.B.Sabin - Bologna - Italy (retired)**
**davideneri100@gmail.com**

**Abstract** - In the Tuscan Archipelago (Italy) it's possible to make an approximate measurement of the Earth's radius from the Elba Island using the Pianosa Island as a ship that appears and disappears on the horizon depending on the observer's height above sea level. The necessary calculations require the mathematical skills of the first two years of high school. The measurement described here refers to a particular geographical location, but it can probably be repeated in similar situations around the world.

**1. A bit of history**
Greek astronomers and geographers discovered the spherical shape of Earth 2,500 years ago. In ancient times, two different values for the circumference were proposed: one by Eratosthenes ($3^{rd}$ century BC: 252,000 stadia) and the other by Ptolemy ($2^{nd}$ century AD: 180,000 stadia). The measurements were made by observing the Sun or stars from different locations. Since the stadium, in the Hellenistic period, corresponded to 155-185 metres, Eratosthenes value seems to be the most accurate.
Ancient geographers were also aware that the sphere is only an approximation of the shape of the Earth. In his treatise *The heavens*, Cleomedes writes: "Those who say that the Earth cannot be spherical because of the hollows occupied by the sea and the mountainous protrusions, express a quite irrational doctrine. […] The protrusions on the rondures from plane trees also do not stop them from being rondures. Yet these protrusions have a ratio to the total sizes of the rondures greater than that of the hollows of the sea and the mountainous protrusions to the total size of the Earth." [1]
Contrary to a widespread belief, although in the early Middle Ages in Western Europe many scientific knowledge has been forgotten, that of the sphericity of the Earth has never been interrupted because some late ancient encyclopedic works, such as the *Commentary on the Dream of Scipio* by Macrobius and *On the Marriage of Philology and Mercury* by Martianus Capella, have handed it down without interruption up to the present day. The latter, in particular, explicitly cites the values that Eratosthenes and Ptolemy attributed to the circumference of the Earth (Book VI, §§ 596-598 and 609-610). Starting from the $12^{th}$ century, thanks to the translation in Latin of ancient works from Greek and Arabic, texts that talk about the spherical Earth became more numerous. Even non-specialist authors talk about it, such as Thomas Aquinas (in the first pages of the *Summa theologica*) and Dante Alighieri (*Divina Commedia*, Inferno XXVI, and in the *Convivio*, where a value of Earth circumferencce obtained from Arabic sources is provided). Another prejudice that is still widely accepted concerns the geographers who, at the end of the $15^{th}$ century, criticized Columbus' idea of reaching Asia by sailing towards the west: they didn't believe in a flat Earth, but they relied on Eratosthenes and rightly affirmed that Japan was too distant to be reached by caravels. Columbus, who followed Ptolemy's assessment, believed the Earth was smaller than it is, and he was lucky enough to find unknown lands halfway on his journey road. An image of the Earth close to Ptolemy's ideas is represented by Behaim's globe (about 1490) still preserved in Nuremberg.

**2. The proof of the ship**
One of the best known proofs in favor of the sphericity of the Earth can be obtained by observing a ship moving away from the coast: it is known that the hull disappears while the mast is still visible, even if the distance would be such as to allow you to still see the whole ship. The reason is in the curvature of the Earth, which only covers the lower part of the ship.

The topic has been widely used since ancient times: it is found for example in the *Natural History* of Pliny the Elder. In the *Almagest*, Ptolemy instead mentions the heights of the mainland which, when one ship approaches the coast, suddenly appear emerging from water whereas, if the Earth were flat, it would have been possible see them from a greater distance. The topic is taken up again in one of the most widespread medieval treatises, *The Sphere* of John of Holywood (about 1250), where it is often accompanied by illustrations such as that of figure 1.

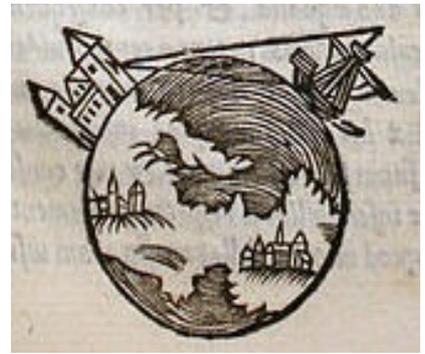

*Fig. 1*

A similar observation can be made in the Northern Tyrrhenian Sea, by looking at the Pianosa island, a rock platform which is located south-west of the Elba island and behaves like a stationary ship in the sea. Pianosa also has a "mast" visible from afar at night: the lantern of its lighthouse. From the observation of the lighthouse at suitable times and places, combined with geometry and algebra taught in the high school, the radius and the circumference of the Earth can be obtained with good approximation. The fact that Elba and Pianosa are at a fixed distance allows us to repeat the observations at will.

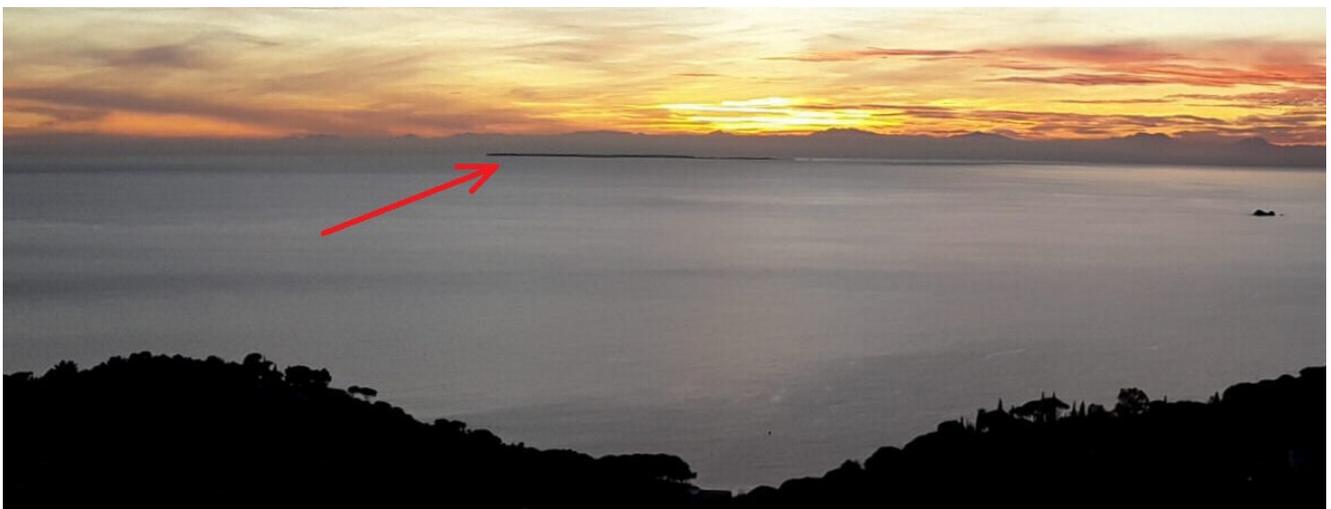

*Fig. 2 - The thin dark stripe indicated by the red arrow is Pianosa Island seen from Capoliveri. Madonna delle Grazie beach is in the cove at the bottom center of the photo.*

**3. A few simple observations**

Let's start with some easy-to-find data:

1] the town of Capoliveri, on the Elba Island, is about 160 meters above sea level, and the island of Pianosa can be seen from the town distinctly (fig.2). The beach of Madonna delle Grazie, located near Capoliveri in the direction of Pianosa is, obviously, a few centimeters above sea level and, during the day, from the beach Pianosa is not visible. This is enough to disprove the flat Earth theory.

2] the distance between the beach and the lighthouse, which can be obtained approximately from a road atlas, corresponds to (28.0±0.2) km (fig. 3). The same evaluation can be obtained using the tools made available by Google Maps. In the following calculations the value $d$=28 km will therefore be used.

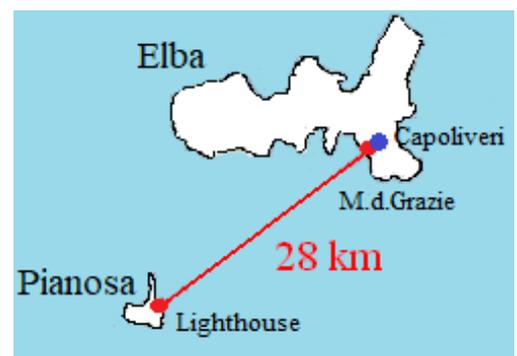

*Fig. 3*

3] the lantern of Pianosa lighthouse is 42 meters above sea level [2]. From Madonna delle Grazie beach, the Pianosa lighthouse is visible only if the eye of the beholder is not too close to sea level. Below a certain height $X$ the lighthouse can no longer be seen. The value of $X$ is not easy to determine, because sea level varies continuously. Repeated observations, accomplished by avoiding operating in extreme tidal conditions, indicate only that, in Madonna delle Grazie beach, $X$ is approximately between 1.60 and 2.00 metres. The evaluation of $X$, however approximate, allows to estimate the Earth's radius.

The difficulty of measuring $X$ can be explained by the continuous variation of sea level due to astronomical and meteorological phenomena. The astronomical tide is produced by the gravitational force of the Sun and Moon and is calculated with remarkable precision. The meteorological tide is instead due to atmospheric pressure and to the wind. Obviously, the meteorological tide is more irregular and unpredictable than the astronomical tide.

These phenomena change the height of the lighthouse above sea level and affect the measure of $X$, which should refer to the medium sea level. To give an idea of the extent of these effects in the Northern Tyrrhenian Sea, I report the values referring to Piombino [3], which is located approximately 30 km north-east of the sea area between Elba and Pianosa: the astronomical tide can produce maximum oscillations between those of Livorno (±18 cm) and Civitavecchia (±20 cm), the effect of atmospheric pressure varies between +38 cm and −31 cm. The wind, when it blows towards the coast, can raise sea level up to 20 cm. To minimize errors it is therefore advisable to take measurements in the absence of wind, when the atmospheric pressure approaches the normal value of 1013 mb, and to consult the charts of astronomical tides available on the Web for many coastal locations.

### 4. A bit of geometry

Figure 4 summarizes the data just exposed. Obviously the scale is not in scale: the Earth's radius $R$ is very underestimated and the curvature of the earth is greatly amplified.

If below a height $X$ of the observer's eye the Pianosa lighthouse disappears, there must be a point P on the surface of the sea where the observation line FO is tangent to the Earth's circumference. Then, according to Euclidean geometry, $\angle CPF = \angle CPO = 90°$, and $\triangle CFP$ and $\triangle CPO$ are right triangles.

To calculate $R$ we must determine the lenght of $\overline{FP}=y$ in the first triangle, $\overline{PO}=z$ in the second, knowing that $\overline{FO}=y+z \approx 28$ km. The size of the Earth's radius, definitely not small, allows us to consider equivalent the segment $\overline{FO}$ and the circumferential arc included between the base of the lighthouse and the observer.

### 5. A bit of algebra

The calculations necessary to obtain the Earth's radius R based on the height $X$ can be done by any student who knows the Pythagorean theorem and the fundamental rules of literal calculus.

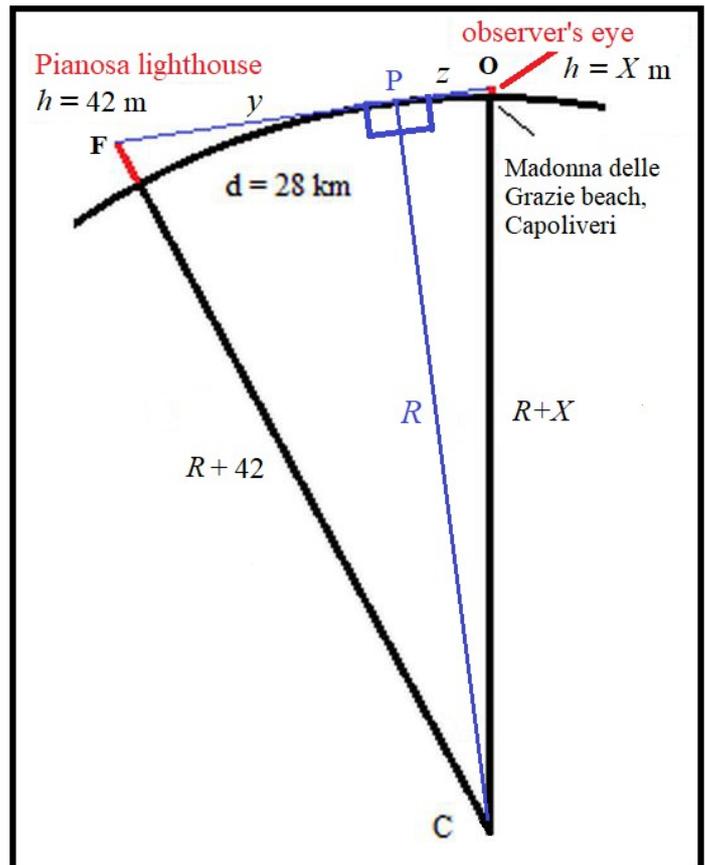

*Fig. 4*

These calculations are developed in this paragraph and simplified by introducing some approximations.

Applying the Pythagorean theorem to the two right triangles CFP and CPO we obtain, respectively:

$$R^2 + y^2 = (R+42)^2, \qquad (1)$$
$$R^2 + z^2 = (R+X)^2. \qquad (2)$$

from which we have:

$$(R+42)^2 - y^2 = (R+X)^2 - z^2. \qquad (3)$$

From this equation we obtain the difference of squares $y^2 - z^2 = (84-2X)R+(1764-X^2)$. Being $y+z \approx 28$ km = 28000m, we find that $z-y = (y^2 - z^2)/28000$. The term $(1764-X^2)$, when it comes divided by 28000, is negligible and can be ignored. It is thus obtained that, in meters,

$$y \approx 14000 + [(42-X)R]/28000 \qquad (4)$$
$$z \approx 14000 - [(42-X)R]/28000 \qquad (5)$$

We can insert these expressions into equations (1) and (2) to calculate $R$ for the two values $X$=2.00 m and $X$=1.60 m, that determine the limits of observations.

Case *X* =2.00 m.
With a few algebraic steps we obtain by eq. (1) and (2), respectively:
$$(R/700)^2 − 44R + 196{,}000{,}000 − 1764 = 0, \quad (6)$$
$$(R/700)^2 − 44R + 196{,}000{,}000 − 4 = 0. \quad (7)$$
The value 1764 in eq. (6) and the value 4 in eq. (7) can be neglected because their are very small compared to 196,000,000. After this simplification the two equations coincide and become a single second degree equation, with *R* in meters:
$$(R/700)^2 − 44R + 196{,}000{,}000 \approx 0,$$
that we can write
$$R^2 − 2.156 \times 10^7 R + 9.604 \times 10^{13} \approx 0. \quad (8)$$
The equation has two solutions: $R \approx 6290$ km and $R \approx 15270$ km. With $R \approx 6290$ km, for the eqs. (4) and (5) we have $y \approx 23.1$ km, $z \approx 4.9$ km. This solution is therefore acceptable. With $R \approx 15260$ km we instead have $y \approx 35.8$ km, $z \approx −7.8$ km. This second solution is not acceptable because *z*, which represents a length, cannot be negative.

Case *X* =1.60 m.
Proceeding as in the previous case we obtain by eq. (1) and (2), respectively:
$$(R/693)^2 − (43.6)R + 196{,}000{,}000 − 1764 \approx 0, \quad (9)$$
$$(R/693)^2 − (43.6)R + 196{,}000{,}000 − 2.56 \approx 0. \quad (10)$$
As in the previous case, the values 1764 in eq. (9) and 2.56 in eq. (10) are negligible compared to 196,000,000. One thus obtains a single second degree equation, with *R* in meters:
$$(R/693)^2 − (43.6)R + 196{,}000{,}000 \approx 0$$
and therefore (with small roundings)
$$R^2 − 2.094 \times 10^7 R + 9.413 \times 10^{13} \approx 0. \quad (11)$$
The solutions are: $R \approx 6535$ km and $R \approx 14405$ km. With $R \approx 6535$ km, from eqs. (4) and (5) we have $y \approx 23.4$ km, $z \approx 4.6$ km. The solution is therefore acceptable. From $R \approx 14405$ km we obtain $y \approx 34.8$ km, $z \approx −6.8$ km. As in the previous case, this second solution is not acceptable because *z* cannot be negative.

**Conclusion**
In summary, two extreme values of *R* were deduced from the extreme values of *X* obtained from the observations: when *X*=2.00 m we have $R \approx 6290$ km; when *X*=1.60 m we have $R \approx 6535$ km.
The currently known mean radius of the Earth $R_M$ is about 6371 km and lies between the two extremes obtained by the measurements. The Earth's circumference that can be obtained from the extreme values of the radius is then between 39520 and 41090 km, while the one calculated using the average accredited mean radius is approximately 40030 km.
Obviously, the uncertainty associated with the measurement does not allow us to make assessments on the difference among the model of the spherical Earth, the reference ellipsoid and the geoid that defines the mean sea level. One can only note that, in the area considered (Northern Tyrrhenian Sea, with latitude ≈42.7°N), $R_M$ is almost equal to the corresponding value (≈ 6370 km) obtainable from the Earth ellipsoid with semi-major and semi-minor axes $a \approx 6378$ km and $b \approx 6357$ km, and to the value of the geoid, which is only 50 m higher than the ellipsoid [4].

In any case, details about the exact shape of the Earth concern people who trust the scientific method. A simple and approximate measure like the one proposed here should help those who are not yet fully aware of it. If similar measurements can be carried out in other places in the world, they can be very interesting from an educational point of view.